\documentclass[12pt]{article}
\usepackage{amsmath}
\usepackage{amsfonts}     
\usepackage{dsfont}
\usepackage{graphicx,psfrag,epsf}
\usepackage{enumerate}
\usepackage{url} 
\usepackage{multirow}
\usepackage{float}
\restylefloat{table}
\usepackage{graphicx}
\usepackage{pdflscape}
\usepackage{listings}
\usepackage{natbib}
\floatstyle{plaintop}
\restylefloat{table}
\usepackage{dcolumn}
\usepackage{multirow}
\usepackage{caption} 
\newcommand{\blind}{0}

\newcommand{\bo}{\boldsymbol{O}}
\newcommand{\bg}{\boldsymbol{\gamma}}
\bibpunct{(}{)}{;}{a}{}{,}

\begin{document}

\def\spacingset#1{\renewcommand{\baselinestretch}%
{#1}\small\normalsize} \spacingset{1}

\if0\blind
{
  \title{\bf On Inverse Probability Weighting for Nonmonotone Missing at Random Data}
  \author{BaoLuo Sun \hspace{.2cm}\\
    Department of Biostatistics, Harvard School of Public Health\\
    and \\
    Eric J. Tchetgen Tchetgen \\
    Departments of Biostatistics \& Epidemiology, Harvard School of Public Health}
    \date{}
  \maketitle
} \fi

\if1\blind
{
  \bigskip
  \bigskip
  \bigskip
  \begin{center}
    {\LARGE\bf Title}
\end{center}
  \medskip
} \fi

\bigskip
\begin{abstract}
The development of coherent missing data models to account for nonmonotone missing at random (MAR) data by inverse probability weighting (IPW) remains to date largely unresolved. As a consequence, IPW has essentially been restricted for use only in monotone missing data settings. We propose a class of models for nonmonotone missing data mechanisms that spans the MAR model, while allowing the underlying full data law to remain unrestricted. For parametric specifications within the proposed class, we introduce an unconstrained maximum likelihood estimator for estimating the missing data probabilities which can be easily implemented using existing software. To circumvent potential convergence issues with this procedure, we also introduce a Bayesian constrained approach to estimate the missing data process which is guaranteed to yield inferences that respect all model restrictions. The efficiency of the standard IPW estimator is improved by incorporating information from incomplete cases through an augmented estimating equation which is optimal within a large class of estimating equations. We investigate the finite-sample properties of the proposed estimators in a simulation study and  illustrate the new methodology in an application evaluating key correlates of preterm delivery for infants born to HIV infected mothers in Botswana, Africa. 

\end{abstract}

\noindent%
\vfill

\newpage
\spacingset{1.45} 
\section{INTRODUCTION}
\label{sec:intro}

Missing data is a major complication which occurs frequently in empirical research. Non-response in sample surveys, dropout or non-compliance in clinical trials and data excision by error or to protect confidentiality are but a few examples of ways in which full data is unavailable and our ability to make accurate inferences may be compromised. Missingness could also be introduced into a study by design, e.g. multi-stage sampling plans in order to reduce the cost associated with measurements for all subjects. In many practical situations, the missing data pattern is non-monotone, that is, there is no nested pattern of missingness such that observing variable $X_k$ implies that variable $X_j$ is also observed, for any $j <k$. Non-monotone missing data patterns may occur, for instance, when individuals who dropped out of a longitudinal study re-enter at later time points. The missing data process is said to be missing-completely-at-random (MCAR) if it is independent of both observed and unobserved variables in the full data, and missing-at-random (MAR) if, conditional on the observed variables, the process is independent of the unobserved ones \citep{rubin1}. A missing data process which is neither MCAR nor MAR is said to be missing-not-at-random (MNAR). 

While complete-case (CC) analysis is the easiest to implement and often employed in practice, the method is generally known to produce biased estimates when the missingness mechanism is not MCAR \citep{rubin1}, although in regression settings, a CC analysis remains unbiased provided the missingness process does not depend on the outcome given observed covariates included in the regression model \citep{rubin1, littlez}. Other commonly used procedures include last-observation-carried-forward analysis most commonly used in longitudinal studies and other single imputation techniques. However, such ad-hoc approaches typically provide valid inferences only under restrictive and often unrealistic conditions \citep{molen1, Siddi1, rubin1}. The development of principled methods to appropriately account for missing data has been an area of active and on-going research. The assumptions of MAR or MCAR, together with separability of parameters governing the missingness mechanism and complete data model, provide sufficient conditions for valid inferences based on the observed data likelihood \citep{rubin1}. Multiple Imputation (MI) is also another popular technique to account for missing data since its introduction in the context of survey studies \citep{rubin4}, and is widely utilized through its incorporation into mainstream statistical software \citep{horton}. 

Inverse probability weighting (IPW) \citep{horvitz,rubin1,jamie2,van,tsia} creates a pseudo-population of complete cases in which selection bias due to missing data is removed. IPW estimation does not require specification of the full-data likelihood, but the missingness mechanism needs to be modeled. The development of coherent models and practical estimation procedures for the missingness probabilities of nonmonotone missing data is challenging, even under the assumption that the data is MAR. To the best of our knowledge, and as discussed in the seminal missing data book of  \citet{tsia}, there currently is not available, a general approach to model an arbitrary nonmonotone missing data generating process strictly imposing MAR only. This represents an important gap in the missing data literature, which has essentially restricted the use of inverse probability weighted estimation to monotone missing data settings.

In this paper, we propose a class of models for arbitrary nonmonotone MAR data patterns. In order to estimate the missingness mechanism required for IPW estimation, we present two approaches: unconstrained maximum likelihood estimation (UMLE) and constrained Bayesian estimation (CBE). The first approach is easily implemented in standard software, say using existing procedures in SAS or R. However, despite this appealing feature, as we illustrate in the simulation studies, UMLE has a major drawback, in that it is not guaranteed to converge in finite sample, even if all regression models are correctly specified. This problematic feature of the approach is mainly due to the fact that it fails to impose certain natural restrictions of the model. In addition to UMLE, we introduce a CBE approach \citep{bayes} which largely resolves any convergence difficulty and is easily implemented in standard Bayesian software packages. As IPW may be inefficient in practice, we improve its asymptotic efficiency by recovering available information from incomplete cases through implementing an augmented IPW (AIPW) estimator which is optimal within a very large class of AIPW estimators. The approach, which combines the proposed estimators of the nonmonotone missing data process with ideas originating from the seminal work of \citet{jamie2} and further developed by \citet{van} and \citet{tsia}, holds appeal in its relative simplicity, and in the fact that it leverages available information from incomplete cases without having to specify a model of the full data distribution. We present a simulation study to investigate the finite-sample properties of both constrained and unconstrained inferences in the context of logistic regression with nonmonotone missing outcome and covariates, followed by an analysis of preterm delivery on a cohort of women in Botswana to illustrate an application of the methods.

\section{NOTATION AND ASSUMPTIONS}
\label{sec:not}

Let $L=(L_1,...,L_K)^\prime$ be a random K-vector representing the complete data. Let $R$ be the scalar random variable encoding the different missing data patterns. For each of $n$ individuals, we observe an independently and identically distributed realization of $\left(R,L_{(R)}\right)$. For missing data pattern $R=m$, where $1\le m\le2^K-1$, we only observe $L_{(m)} \subseteq L$.  We reserve $R=1$ to denote complete cases. Let $\mathbb{P}_n$ denote the empirical measure $\mathbb{P}_n f(O) = n^{-1} \sum_i f(O_i)$.

For non-parametric identification of the missing data model, we assume that the missing data process is MAR \citep{jamie2}
\begin{align} \label{eq:assume1}
\Pr \{R=m|L\}=\Pr \{R=m|L_{(m)}\},\phantom{-}  \phantom{-}  1\le m\le2^K
\end{align}
so that the conditional probability of having missing data pattern $m$, which we denote by $\pi_m \left(L_{(m)}\right)$, depends only on the observed variables for that pattern. Throughout, we also make the following positivity assumption
\begin{align}\label{eq:assume2}
\pi_1 (L)>\sigma >0 \phantom{-} \text{with probability 1},  
\end{align}
for a fixed positive constant $\sigma$, that is, the probability of being a complete case is bounded away from zero. Assumption (\ref{eq:assume2}) is necessary for identification of the full data law and smooth functionals of the latter \citep{jamie2}, and ensures finite asymptotic variance of the IPW and AIPW estimators. 

A key implication of the MAR assumption is that the missing data process is nonparametrically identified. This also implies that if separate parameters index the missing data mechanism and the full data distribution, efficient estimation of the parameters of the missing data process can be obtained by maximizing its partial likelihood, ignoring the part of the likelihood corresponding to the full data.

\section{ESTIMATION OF MISSING DATA MECHANISM}

Although the missingness mechanism is in principle nonparametrically identified under assumptions (\ref{eq:assume1}) and (\ref{eq:assume2}), in practice estimation typically entails specifying parametric models as dictated by the curse of dimensionality, since $L$ is typically of moderate to high dimension \citep{rit}. To motivate our discussion of nonmonotone missing data models, we briefly review strategies for modeling some common missing data structures. In the simple case of two missing data patterns, i.e. $R=1,2$, the probability of being a complete case is $1-\pi_2 \left(L_{(2)}\right)$ and the parameters $\gamma$ of a model $\pi_2 \left(L_{(2)};\gamma\right)$ can be estimated by maximizing the likelihood function
$$
\prod_i \left\{ 1-\pi_2 \left(L_{(2)};\gamma\right) \right\}^{\mathds{1}(R_i=1)}\left\{ \pi_2 \left(L_{(2)};\gamma\right)\right\}^{1-\mathds{1}(R_i=1)}.
$$
The two-missing-data-pattern scenario arises in familiar settings such as in regression analysis with incomplete data only on the outcome for a subset of the sample.

When $M>2$ the missing data is said to be monotone if for some ordering of the variables in $L$, the $k^{\text{th}}$ variable is observed only if the $k-1^{\text{th}}$ variable was observed, and therefore one can sort the missing data patterns in such a way that $L_{(m+1)} \subset L_{(m)}$ for $m=1,...,M-1$. Then the missing data mechanism can be modeled using a discrete hazard function \citep{jamie2, tsia} by defining
\begin{align*}
\lambda_m(L_{(m)})=\begin{cases}
 \Pr(R=m | R \leq m, L), &  m \neq 1. \\
1, & m = 1.
\end{cases} 
\end{align*}
The discrete hazard $\lambda_m(\cdot)$ is a function of $L_{(m)}$ only since
$$
\frac{\Pr(R=m | L)}{ \Pr(R \leq m| L)}= \frac{\pi_m(L_{(m)})}{1-\sum_{j>m}\pi_j(L_{(j)})}
$$
and $L_{(j)} \subset L_{(m)}$ for all $j>m$ by the monotone missing data structure. Defining
\begin{align*}
K_m(L_{(m)})&= \Pr(R<m | L)=\prod_{j\geq m} \left\{ 1- \lambda_j(L_{(j)}) \right\},\phantom{-}  m \neq 1,
\end{align*}
the conditional probability for each missing data pattern is
\begin{align*}
\pi_m(L_{(m)})=\begin{cases}
K_{m+1}(L_{(m+1)})\lambda_m(L_{(m)}), & m < M.\\
\lambda_m(L_{(m)}), & m = M. 
\end{cases}
\end{align*}
and in particular the complete case probability is
$$
\pi_1(L)=K_2(L_{(2)})= \Pr(R< 2| L)=\prod_{j\geq 2} \left\{ 1- \lambda_j(L_{(j)}) \right\}
$$
To estimate the hazard functions $\lambda_m(L_{(m)})$, in practice we may run a series of logistic regressions of the indicator variable $\mathds{1}(R=m)$ on $L_{(m)}$ among individuals with $R\leq m$, $m=2,...,M$. Alternatively, one may pool information by allowing $\lambda_m(L_{(m)})$ to share parameters across $m$.

\subsection{The failure of standard polytomous regression}
For nonmonotone missing data patterns, the nesting of patterns $L_{(m+1)} \subset L_{(m)}$ is no longer available, and building coherent models for the conditional probabilities of the various missing data patterns is challenging even under assumptions (\ref{eq:assume1}) and (\ref{eq:assume2}) \citep{jamie2, jamie1,  tsia}. A straightforward approach to model $\pi_m(L_{(m)})$ using standard polytomous regression for the multinomial missing data process will often have the unintended consequence of imposing more restrictive  conditions than what MAR assumption (\ref{eq:assume1}) strictly entails \citep{jamie1}, which we illustrate using an example of a general bivariate pattern \citep[pp. 18-19]{rubin1}. Suppose the full data is bivariate $L=(L_1,L_2)$ and one encodes the missing data patterns as follows: $R=1$ if $L$ is observed;  $R=2$ if one only observes $L_{(2)}=L_1$;  $R=3$ if one only observes $L_{(3)}=L_2$; and $R=4$ if neither variable is observed. A standard polytomous logistic regression for $R$ corresponds to
\begin{align} \label{eq:bivar}
\Pr \{R=m|L\}= \frac{\exp(\gamma_{0m}+\gamma_{1m} L_1+ \gamma_{2m} L_2)}{1+\sum_{k=2}^{4} \exp(\gamma_{0k}+\gamma_{1k} L_1+ \gamma_{2k} L_2) }, \phantom{-}   m=2,3,4.
\end{align}
By the MAR assumption, since for $R=4$ neither variable is observed, the probability $\Pr \{R=4|L\}$ depends on neither $L_1$ nor $L_2$ so that $\gamma_{1j}= \gamma_{2j}=0$ for $j=2,3,4$. Therefore assuming model (\ref{eq:bivar}) under MAR implies MCAR. In general, it is shown in appendix A.1 using a similar argument that the missing data pattern probabilities modeled using polytomous logistic regression can at most depend on the intersection of the sets of observed variables $L_{(m)}$, $m=2,3,...,M$, which is strictly stronger than the MAR assumption  (\ref{eq:assume1}). This suggests that standard polytomous regression is ill-suited as modeling strategy for nonmonotone missing data process under MAR.

As a remedy, Robins and Gill proposed a large class of models for the missing data mechanism, which they call the randomised monotone missingness (RMM) processes, that are guaranteed to be MAR for a non-monotone missing data mechanism without necessarily being MCAR \citep{jamie1}. This class of models does not span the space of all MAR models and therefore it is indeed possible to test whether the proposed class of models includes the true missing data mechanism. However, estimation of the missing data mechanism within this class is complex and computationally demanding, even for small to moderate sample size and number of different missing data patterns, and no software is currently available to implement the approach, which has limited its widespread adoption.

\subsection{Proposed nonmonotone missing data model}
Our approach involves modelling the conditional probability for each missing data pattern separately as
\begin{align}  \label{eq:npmiss}
\Pr \{R=m|L\}=\pi_m (L_{(m)}), \phantom{-}  m=2,...,M.
\end{align}
The probability of observing complete data is 
\begin{align}\label{eq:com}
\Pr \{R=1|L\}=\pi_1 (L) = 1 - \sum_{m=2}^{M}\pi_m (L_{(m)}),
\end{align}
which depends on the union set of observed variables $\bigcup_{m=2}^M L_{(m)}$. To ground ideas, consider as a parametric submodel of (\ref{eq:npmiss}) the series of simple logistic models 
\begin{align}  \label{eq:logit}
\notag \pi_m \left(L_{(m)} ; \gamma_m\right) &=\left\{ 1+ \exp \left[ -{\gamma_m} \left(1,L_{(m)}\right)^T \right]\right\}^{-1}, \phantom{-}  m=2,...,M, \\ 
\pi_1 (L ; \gamma) &= 1 - \sum_{m=2}^{M}\left\{ 1+ \exp \left[ -{\gamma_m} \left(1,L_{(m)}\right)^T \right]\right\}^{-1}, \phantom{-} \gamma=(\gamma_2,...,\gamma_M).
\end{align}
By assumption (\ref{eq:assume2}), model (\ref{eq:logit}) must satisfy the constraint
\begin{align}\label{eq:cons}
1-\sum_{m=2}^{M}  \pi_m \left(L_{(m)} ; \gamma_m\right) > \sigma \phantom{-} \text{          with probability 1}. 
\end{align}
Consider the UMLE estimator of $\gamma$, defined as the value which maximizes the unconstrained log-likelihood function corresponding to missing data model (\ref{eq:logit}).
 \begin{align}\label{eq:loglike}
\sum_{i=1}^{N} \left\{ \left[  \sum_{m=2}^{M} \mathds{1}(R_i=m) \log \pi_m \left(L_{(m)i} ; \gamma_m\right) \right] + \mathds{1}(R_i=1) \log \left[ 1 - \sum_{k=2}^{M} \pi_k \left(L_{(k)i} ; \gamma_k\right) \right] \right\}
\end{align}
with corresponding score equation
\begin{align}\label{eq:scores}
\mathbb{P}_n \left\{ \left[ \frac{\mathds{1}(R=1)}{\pi_1 \left(L_{(1)}\right)}-\frac{\mathds{1}(R=m)}{\pi_m \left(L_{(m)}\right)}\right]\pi_m (1-\pi_m ) \left(1, L_{(m)}\right)^T \right\} =0
\end{align}
for the parameters $\gamma_m$ for missing data pattern $m$, where $\gamma_m$ and $\left(1, L_{(m)}\right)^T$ have the same dimension. 

It may be in practice that maximizing (\ref{eq:loglike}) fails to converge. This could happen if there is at least one individual for whom the empirical version of constraint (\ref{eq:cons}) is not satisfied in the process of finding the maximum, in which case the fitted complete case probability may be near zero or possibly negative, a real possibility especially at small or moderate sample sizes. Thus, we have referred to (\ref{eq:loglike}) as an unconstrained log-likelihood function, as it does not naturally impose constraint (\ref{eq:cons}).

Note that even if the missingness mechanism were known, constraint (\ref{eq:cons}) which depends on $\bigcup_{m=2}^M L_{(m)}$ can only be observed for complete case individuals. In fact, only complete cases need to satisfy the constraint in order to ensure that the UMLE can be computed in practice. Thus, one could in principle attempt to maximize the observed data log-likelihood (\ref{eq:loglike}) together with the observable constraints
 \begin{align}\label{eq:cons2}
 \mathds{1}(R_i=1)\sum_{k=2}^{M}  \pi_k \left(L_{(k)i} ; \gamma_k\right) <1-\sigma^* \phantom{-} \text{       for } i=1,2,...,N,
\end{align}
where $\sigma^*$ is a user-specified small positive constant. Still, this is potentially computationally prohibitive, since there are as many constraints as complete case observations. 

Instead, in addition to UMLE, we develop a Bayesian constrained estimation approach where samples are drawn from the unconstrained posterior conditional distribution for $\gamma$ and only those draws that fall into the constrained parameter space (\ref{eq:cons2}) are retained \citep{bayes}. An additional appeal of this approach is that the posterior credible intervals of $\gamma$ are guaranteed to satisfy constraint (\ref{eq:cons2}), which is useful if one wishes to perform hypothesis testing to identify significant predictors in the missing data regression models. Constrained Bayesian estimation has been used previously in several other settings, for instance to estimate risk ratio and relative excess risk regressions \citep{chucole1, chucole2}; however, to the best of our knowledge, it has not been used in the current context. To implement the approach, we specify a diffuse prior distribution $g(\gamma)$ for $\gamma=(\gamma_2,...,\gamma_M)$ under model (\ref{eq:logit}) and incorporate constraint (\ref{eq:cons2}) in the posterior distribution of $\gamma$. Under the contrained Bayesian model, the posterior distribution of $\gamma$ is proportional to
\begin{align}\label{eq:pos}
f(\gamma | data) \propto f(data|\gamma) g(\gamma) = \prod_{i=1}^{N} \left\{ \prod_{m=2}^{M} \left\{ \pi_{m} \left(L_{(m)i} ; \gamma_{m}\right) \right\}^{ \mathds{1}(R_i=m)}  \times \Omega(\gamma,L_i)^{ \mathds{1}(R_i=1)} \vphantom{\frac12}\right\}  g(\gamma) 
\end{align}
where 
$$
 \Omega(\gamma,L_i)= \left\{ \left[1-\sum_{k=2}^{M}  \pi_{k} \left(L_{(k)i} ; \gamma_{k}\right) \right]\times   \mathds{1}\left[\sum_{k=2}^{M}  \pi_{k} \left(L_{(k)i} ; \gamma_{k} \right) <1-\sigma^* \right]\right\}.
$$
We define the CBE estimator of $\gamma$ as the posterior mode (or mean) from distribution (\ref{eq:pos}). 

We note that in practice there may be some missing data patterns that are sparsely observed. In such cases, a simple approach entails combining across patterns with small event probabilities and estimating the missingness process under an additional assumption that the probability of any pattern within the combined set only depends on the intersection set of variables observed for all patterns in the combined set. Although the suggested approach to handle sparse patterns may introduce some bias, we do not anticipate the magnitude of this bias to be significant provided the combined set of patterns remains relatively rare compared to other more prominent missing data patterns.

\section{IPW INFERENCE}
\label{sec:ipw}

Suppose we observe $n$ i.i.d. realizations of the vector $L$, and we wish to make inferences about the parameter $\beta_0$ which is the unique solution of the full data population estimating equation
\begin{align}\label{eq:nomiss}
E\{M(L;\beta_0)\}=0 
\end{align}
where expectation is taken over the distribution of the complete data $L$. Note that we do not require a model for the distribution of the full data $L$; in fact, estimation is possible under certain weak regularity conditions \citep{vaart} as long as full data unbiased estimating functions exist. In the presence of missing data, the estimating function in (\ref{eq:nomiss}) may only be evaluated for complete cases, which may be a highly selective subsample even under MAR. This motivates the use of IPW estimating functions of complete cases to form the following population estimating equation
\begin{align}\label{eq:ipw}
E\left\{\frac{\mathds{1}(R=1)}{ \pi_1 (L)}M(L;\beta_0)\right\}=0.
\end{align}
The unbiasedness of the above estimating equation holds by straightforward iterated expectations.

The IPW estimating equations framework encompasses a great variety of settings under which investigators may wish to account for non-monotone missing data. This includes IPW of the full data score equation, where the score function is such an unbiased estimating function, given a model $f(L | \beta)$ for the law of the full data, in which case (\ref{eq:ipw}) reduces to
\begin{align}\label{eq:ipwscore}
E\left\{\frac{\mathds{1}(R=1)}{ \pi_1 (L)}\frac{\partial \log f(L|\beta)}{\partial \beta} \Bigg|_{\beta_0}\right\}=0.
\end{align}

Note that equation (\ref{eq:ipwscore}) does not necessarily correspond to the observed data score equation, and will therefore generally not achieve the efficiency bound for the model. Estimation can also be extended to classes of semiparametric models which specify only certain marginal relationships in $L$ and in which scientific interest focuses on some low dimentional functional $\beta=\beta(F_{L})$ of the distribution $F_{L}$ of the full data $L$. For instance, in many health related applications it is common to specify a model $g(X,\beta)$ for the conditional mean of the outcome response $Y$ given a set of covariates $X=(X_1,X_2,...,X_P)^T$. Here $L=(Y,X)$ and either the outcome or any covariate may be missing. Then the parameter of interest can be identified by the population IPW estimating equation
\begin{align*}
E\left\{\frac{\mathds{1}(R=1)}{ \pi_1 (L)}\left[Y-g(X,\beta_0)\right]h(X)\right\}=0,
\end{align*}
where $h(X)$ is a user-specified function of $X$ of the same dimension as $\beta_0$. Regression parameters in semiparametric models for right censored failure time data can likewise be identified by similar IPW population estimating equations, e.g. Cox proportional hazards regression and Aalen's additive hazards regression. Analogous estimating equations are also available for longitudinal and clustered data. In all cases empirical estimating equations are obtained by replacing population expectations with their empirical counterparts, and $\pi_1 (L)$ with a consistent estimator. 

Let ${\pi}_1 (L;\hat{\gamma}) = 1 - \sum_{m=2}^{M}\left\{ 1+ \exp \left[ -{\hat{\gamma}_m} \left(1,L_{(m)}\right)^T \right]\right\}^{-1} $ where $\hat{\gamma}=(\hat{\gamma}_2,...\hat{\gamma}_M)$ is either the UMLE (assuming it can be computed) or CBE estimate. Then, an estimate for the parameter of interest $\beta_0$ is given by the solution $\hat{\beta}_{ipw}$ to the inverse probability weighted estimating equation
\begin{align} \label{eq:ipweqn}
\mathbb{P}_n \left\{\frac{ \mathds{1}(R=1)}{ \pi_1 (L;\hat{\gamma})}M(L;\beta)\right\}=0.
\end{align}
Subject to standard regularity conditions and assuming that the missing data model given in (\ref{eq:logit}) is correctly specified, we show in appendix section A.2 that $\hat{\beta}_{ipw}$ is consistent and asymptotically normal
\begin{align}\label{eq:ipwavar}
\sqrt{n}(\hat{\beta}_{ipw}-\beta_0) \xrightarrow[]{d} N\left(0, E\{\nabla_\beta \Gamma(\beta_0,\gamma_0) \}^{-1}\mathrm{Var}\left[  \Gamma(\beta_0,\gamma_0)-W(\beta_0,\gamma_0)\right]E\{\nabla_\beta \Gamma(\beta_0,\gamma_0) \}^{-1^T}\right)
\end{align}
 where $ \Gamma(\beta,\gamma) = \{ \mathds{1}(R=1)/\pi_1 (L;\gamma)\}M(L;\beta)$, $S_{\gamma_0}$ is the score function (\ref{eq:scores}) for the missing data mechanism evaluated at the truth and 
$$
W(\beta_0,\gamma_0)=E \left[\Gamma (\beta_0,\gamma_0)S_{\gamma_0}^T\right]E \left[S_{\gamma_0} S_{\gamma_0}^{T}\right]^{-1} S{\gamma_0}.
$$
The asymptotic variance in (\ref{eq:ipwavar}) can be consistently estimated by replacing the terms under expectation with empirical averages evaluated at $\left(\hat{\beta}_{ipw},\hat{\gamma}\right)$
\begin{align}
\widehat{E}\{\nabla_\beta \Gamma(\hat{\beta},\hat{\gamma}) \}^{-1}\widehat{\mathrm{Var}}\left[  \Gamma(\hat{\beta},\hat{\gamma})-\widehat{W}(\hat{\beta},\hat{\gamma})\right]\widehat{E}\{\nabla_\beta \Gamma(\hat{\beta},\hat{\gamma}) \}^{-1^T}.
\end{align}
Although the posterior mode (or mean) is asymptotically efficient by the Bernstein-von Mises Theorem \citep{vaart}, in finite sample the BCE estimate may not necessarily correspond to the solution of the score function (\ref{eq:scores}). For inference under the Bayesian constrained approach, we therefore apply a finite-sample correction to the variance estimate 
\begin{align}
\widehat{{E}}\{\nabla_\beta \Gamma(\hat{\beta},\hat{\gamma}) \}^{-1}\widehat{\mathrm{Var}}\left[ \Gamma(\hat{\beta},\hat{\gamma})-\widehat{W}(\hat{\beta},\hat{\gamma})+\widehat{{E}}\{W(\hat{\beta},\hat{\gamma})\} \right]\widehat{{E}}\{\nabla_\beta \Gamma(\hat{\beta},\hat{\gamma}) \}^{-1^T}
\end{align}
so that the term in $\widehat{\mathrm{Var}}[\cdot]$ has mean zero empirically. The correction term $\widehat{E}\{W(\hat{\beta},\hat{\gamma})\}$ is expected to vanish as sample size increases. A conservative, albeit more easily implementable, estimate of the asymptotic variance in (\ref{eq:ipwavar}) is obtained by the standard sandwich variance formula \citep{jamie2}
\begin{align}
\widehat{{E}}\{\nabla_\beta \Gamma(\hat{\beta},\hat{\gamma}) \}^{-1}\widehat{\mathrm{Var}}\left[  \Gamma(\hat{\beta},\hat{\gamma})\right]\widehat{{E}}\{\nabla_\beta \Gamma(\hat{\beta},\hat{\gamma}) \}^{-1^T}.
\end{align}

\subsection{Improved IPW estimator via augmentation}
The efficiency of the IPW estimator introduced in the previous section, which only makes direct use of complete cases, can be improved by incorporating information from individuals with missing data via augmentation of the IPW estimating equation \citep{jamie2, van, tsia}. The approach is based on a result due to \citet{jamie2} who show that under assumptions (1) and (2), all regular and asymptotically linear (RAL) estimators based on observed data, of a functional $\beta_0$, can be shown to be asymptotically equivalent to an estimator solving
\begin{align} \label{eq:aipw}
\mathbb{P}_n \left\{ \frac{\mathds{1}(R=1)}{\pi_1(L)}U(L;\beta)+A\left(R,L_{(R)}\right) \right\}= 0.
\end{align}
$U(L;\beta)$ is an element of $\mathbb{U}^F$, the set of all full data estimating equations of $\beta_0$, and $A\left(R,L_{(R)}\right)$ is an element of the space $\mathbb{A}$ spanned by all scores of the missing data mechanism which are of the form
\begin{align*}
\left\{ \sum_{r \ne 1} \left[ \frac{\mathds{1}(R=1)}{\pi_1(L)}- \frac{\mathds{1}(R=r)}{\pi_r\left(L_{(r)}\right)}\right]t_r\left(L_{(r)}\right)  \right\},
\end{align*}
where $t_r\left(L_{(r)}\right) $ is an arbitrary $q$-dimensional function of the observed data $L_{(r)}$ corresponding to missing data pattern $R=r$ \citep{jamie2}. The class of estimating equations obtained by varying $U(L)$ over $\mathbb{U}^F$ and $A\left(R,L_{(R)}\right)$ over $\mathbb{A}$ is referred to as augmented estimating equations, since it entails augmenting a standard IPW estimating equation by an arbituary score function of the missingness process \citep{jamie2,tsia}. In principle, one can therefore construct an efficient estimator by identifying the optimal full data estimating function $U_{\text{opt}} \in \mathbb{U}^F$ paired with the optimal choice of augmentation $A_{\text{opt}} \in \mathbb{A}$ to use in equation (\ref{eq:aipw}). Unfortunately the optimal index leading to a semiparametric efficient estimator is generally not available in closed form and often computationally prohibitive in most problems of interest. Instead, we take a more practical approach to improve efficiency.

 We consider the restricted augmentation space $\mathbb{A}^{\ast}\subset\mathbb{A} $ formed by the span of a finite vector of linearly independent functions 
\begin{align*}
\left\{ \left[ \frac{\mathds{1}(R=1)}{\pi_1(L)}- \frac{\mathds{1}(R=r)}{\pi_r\left(L_{(r)}\right)}\right]t_{rk}^{\ast}\left(L_{(r)}\right): r; k=1,...,K_r\right\},
\end{align*}
where for each $r$, $t_{r}^{\ast}(L_{(r)}) $ is a $K_r$-vector of user defined functions of $L_{(r)}$, $r=1,...,M$. It is recommended to include in $\mathbb{A}^{\ast}$ scores corresponding to the model used to estimate the missing data mechanism, which leads to simplification in estimating the asymptotic variance of the resulting estimator \citep{jamie2, tsia}. Specifically, under model (\ref{eq:logit}), $\mathbb{A}^{\ast}$ includes the score functions given by (\ref{eq:scores}).

Similarly, we consider a restricted linear subspace $\mathbb{U}^{F*} \subset \mathbb{U}^F$ spanned by $l$ linearly independent full-data estimating equations, where $l>q$. In the case of logistic regression with full data score equation $U(\beta)=(1,X)^T\left\{Y-\text{expit}\left[\beta(1,X)^T\right]\right\}$ with $L=(X,Y)$, we may take $\mathbb{U}^{F*}$ to be the q-dimensional span of estimating functions 
$$
\left[1,X,h(X)\right]^T\left\{Y-\text{expit}\left[\beta(1,X)^T\right]\right\}
$$
for any choice of function $h(X)$ of dimension $l-q$ linearly independent of $(1,X)$, e.g. including nonlinear transformations of $X$ and interaction terms. The resulting class of restricted augmented estimating equations is given by 
\begin{align} \label{eq:raipw}
\mathbb{P}_n \left\{ \frac{\mathds{1}(R=1)}{\pi_1(L;\hat{\gamma})}C_1 U^*(L;\beta)+C_2 A^*(R,L_{(R)};\hat{\gamma}) \right\}= 0
\end{align}
for any choice of constant matrices $C_1$ of dimensions $q \times l$ and $C_2$ of dimensions $q \times k$ where $k=\sum_{r\ge 2}K_r$. $U^*(L;\beta)$ is a $l$-dimensional vector of basis functions spanning $\mathbb{U}^{F*}$ and $ A^*(R,L_{(R)};\gamma)$ is a $k$-dimensional vector of basis functions spanning $\mathbb{A}^{*}$. Using a result due to \citet{tsia} one can show that the optimal choice of $(C_1, C_2)$ within the class (\ref{eq:raipw}) is given by the solution to 
\begin{align*} 
\left[C_1^{opt}, C_2^{opt}\right]
\begin{bmatrix}
  U_{11} & U_{12} \\
  U_{12}^T & U_{22} 
\end{bmatrix}
=\left[H_1, H_2\right]
\end{align*}
where 
\begin{align*} 
U_{11} &= E\left\{ \frac{U^*(\beta)U^*(\beta)^T}{\pi_1(L)}\right\}^{l \times l} \\
U_{12} &= E\left\{ \frac{\mathds{1}(R=1)}{\pi_1(L)}U^*(\beta)A^{*T}\right\} ^{l \times k}\\
U_{22} &= E\left\{A^*A^{*T}\right\} ^{k \times k}\\
H_{1} &= \left(-E\left\{\frac{\partial U^*(\beta)}{\partial \beta}\right\}^T\right) ^{q \times l} \\
H_{2} &= 0 ^{q \times k}
\end{align*}
The matrices $(U_{11},U_{12},H_{1})$ that involve full data $L$ can be estimated from the complete cases only by standard inverse probability weighted empirical averages and the matrix $U_{22}$ by an empirical average of the observed data. Constrained Bayesian estimation of the missing data process involves centering $A^*$ so that it has mean zero empirically. Then the optimal AIPW estimator $\hat{\beta}_{opt}$ in the restricted class of estimating equations is given by the solution to
\begin{align} \label{eq:araipw}
\mathbb{P}_n \left\{ \frac{\mathds{1}(R=1)}{\pi_1(L;\hat{\gamma})}\widehat{C}_1^{opt}(\beta) U^*(L;\beta)+\widehat{C}_2^{opt}(\beta) A^*(R,L_{(R)};\hat{\gamma}) \right\}= 0,
\end{align}
 and a consistent estimator for the asymptotic variance of $\hat{\beta}_{opt}$ is given by 
 \begin{align} \label{eq:araipwvar}
\left\{ \widehat{H}_1(\hat{\beta}_{opt}) \widehat{U}^{11}(\hat{\beta}_{opt}) \widehat{H}_1^T(\hat{\beta}_{opt})  \right\}^{-1}
\end{align} 
where
 \begin{align*}
\widehat{U}^{11}=\left( \widehat{U}_{11}-\widehat{U}_{12}\widehat{U}_{22}^{-1}\widehat{U}_{12}^T  \right)^{-1}.
\end{align*} 
\citep{tsia}.
Finding the solution $\hat{\beta}_{opt}$ to (\ref{eq:araipw}) involves estimating the matrices for each value of $\beta$, which can be computationally intensive. Instead, an estimator asymptotically equivalent to $\hat{\beta}_{opt}$ is obtained by the simple one-step update of a standard IPW estimator $\hat{\beta}_{ipw}$:
 \begin{align} \label{eq:onestep}
 \hat{\beta}_{opt}^{*} =  \hat{\beta}_{ipw} + \widetilde{IF}_{\beta}\left(\hat{\beta}_{ipw}\right)
\end{align}
where 
 \begin{align*} 
 \widetilde{IF}_{\beta}\left(\hat{\beta}_{ipw}\right)= & \left\{-\sum_i \partial \left[  \frac{\mathds{1}(R_i=1)}{\pi_1(L_i;\hat{\gamma})}M (L_i;\hat{\beta}_{ipw}) \right] / \partial \beta   \right\}^{-1} \times \\
 &\left\{\sum_{i} \left[ \frac{\mathds{1}(R_i=1)}{\pi_1(L_i;\hat{\gamma})}\widehat{C}_1^{opt}(\hat{\beta}_{ipw} ) U^*(L_i;\hat{\beta}_{ipw} )+\widehat{C}_2^{opt}(\hat{\beta}_{ipw} ) A^*(R_i,L_{(R),i};\hat{\gamma}) \right] \right\}
\end{align*}
and $\hat{\beta}_{ipw}$ is the standard IPW solution to (\ref{eq:ipweqn}). It is straightforward to show that under standard regularity conditions and in the absence of model misspecification, the influence function of $\hat{\beta}_{opt}^{*}$ is identical to that of $\hat{\beta}_{opt}$ \citep{vaart}. 

The asymptotic efficiency of the optimal restricted AIPW estimator in relation to the semiparametric efficiency bound for a given full data semiparametric model of interest depends on how close the span of $\mathbb{A}^*$ and $\mathbb{U}^{F*}$ is to $\mathbb{A}$ and $\mathbb{U}^{F}$ respectively. One can show that as one suitably enriches the span of $\mathbb{A}^*$ and $\mathbb{U}^{F*}$ with elements of $\mathbb{A}$ and $\mathbb{U}^{F}$ so that the former two vector spaces increasingly become dense in the latter two subspaces respectively, the asymptotic variance of $n^{1/2}\left(\hat{\beta}_{opt}-\beta_0\right)$ nearly attains the semiparametric local efficiency bound for the semiparametric model of the full data and only other restriction that data are MAR \citep{newey2}.

\section{SIMULATION}
\label{sec:sim}
In this section we report a simulation study to investigate the finite-sample properties of the proposed estimators. Full data consists of independent and identically distributed $\boldsymbol{L}=(Y,A,C)$ with exposure $A$, binary outcome $Y$ and confounders $C=(C_1,C_2)$. We generate missing data with 5 possible patterns: $R=1$ if $L$ is observed;  $R=2$ if $L_{(2)}=(Y,A,C_1)$ is observed;  $R=3$ if $L_{(3)}=(Y,A)$ is observed; $R=4$ if $L_{(4)}=(C_1,C_2)$ is observed and $R=5$ if $L_{(5)}=(Y,C_2)$ is observed. 

The vector $(X_1,X_2,X_3)$ is generated from a multivariate standard normal distribution with correlation coefficient $\rho=0.1$ between $X_1$ \& $X_2$ and $\rho=-0.1$ between $X_1$ \& $X_3$. Then we take $A=\Phi(X_1), C_1=\Phi(X_2)$ and $C_2=\Phi(X_3)$ where $\Phi(\cdot)$ is the CDF of the standard normal distribution. Finally, the outcome variable $Y$ is generated as
$$
\text{logit} \Pr(Y=1 | A, C) = \beta_0 + \beta_1 A + \beta_2 C_1 + \beta_3 C_2
$$
where $\beta=(-0.3, -0.4, 0.3, 0.5)$. The missing data model follows (\ref{eq:logit}) and is given by
\begin{align} \label{eq:simlogit}
\notag \Pr \{R=2|L\} &= \text{expit} \{-1.2-1.2 Y -0.6 A -0.3 C_1 \}\\ 
\Pr \{R=3|L\} &= \text{expit} \{-1.0-0.9 Y  -0.8A \}\\ \notag
\Pr \{R=4|L\} &= \text{expit} \{-1.2-0.7 C_1-0.8 C_2 \} \\ \notag
\Pr \{R=5|L\} &= \text{expit} \{-1.1-1.0 Y-0.8 C_2 \},
\end{align}
with the probability of being a complete case $\Pr \{R=1|L\} = 1 - \sum_{m=2}^{5} \Pr \{R=m|L\}$. The missing data process is generated from a multinomial distribution with the above probabilities, and only the corresponding observed data for the sampled pattern contributes to estimation. We perform 1000 replicates each with sample size $n=1000$ or $2000$. Each simulation replicate has approximately $50 \%$ of complete cases.

The parameters $\gamma$ in the missing data model (\ref{eq:simlogit}) are estimated using both the UMLE and BCE. The UMLE estimator of $\gamma$ is implemented using the R function optim with the quasi-Newton method BFGS. As previously discussed, optimization of the unconstrained likelihood is not guaranteed to converge near the boundary values where an observed complete case probability may be close to zero, leading to very large derivatives of the log-likelihood as indicated by the expression for score equation (\ref{eq:scores}). To investigate situations in which the procedure fails to converge, we approximate the UMLE solution with the derivative-free Nelder-Mead method after a maximum number of iterations. We obtain the BCE estimator of $\gamma$ as the posterior mean of distribution (\ref{eq:pos}) with diffuse priors $\gamma_{j} \sim N\left(0,10^3\right)$ for $j=1,...,13$ and $\sigma^*=10^{-8}$. Adaptive Gibbs sampling \citep{gibbs} was implemented through BRugs, the R interface to the OpenBUGS MCMC software \citep{openbugs}. We assessed convergence by visually inspecting the trace plots as well as through the Gelman-Rubin convergence statistic \citep{gelmanrubin}, and included an adaptive phase of $10^4$ iterations out of a total of $2\times10^4$ iterations. 

Simple IPW logistic regression of the form (\ref{eq:ipweqn}) estimates the coefficient $\beta$ of the outcome regression, based on estimated complete case probabilities using either UMLE or BCE. We further implement the optimal, one-step update AIPW estimator (\ref{eq:onestep}) in the restricted finite-dimensional linear subspaces $\mathbb{A}^{*}$ and $\mathbb{U}^{F*}$ spanned by
$$
\left\{ \left[ \frac{\mathds{1}(R=1)}{\pi_1(L)}- \frac{\mathds{1}(R=r)}{\pi_r(L_{(r)})}\right]\pi_r (1-\pi_r ) t_{rk}^*(L_{(r)}): r;k=1,...,k_r \right\}
$$
and
$$
\left\{ h(A,C)^T\left\{Y-\text{expit}\left[\beta(1,A,C)^T\right]\right\} \right\}
$$
respectively. The $1 \times 9$ vector $h(A,C)$ consists of the main effects and interactions for $(A,C)$ up to quadratic terms. The $J_r$-dimensional vectors $t_{r}^*(L_{(r)})$ consists of the $J_r$ main effects and interactions for $L_{(r)}$ up to quadratic terms, for $r=2,3,4,5$. 

For comparison, we also implement unweighted complete-case (CC) regression to evaluate the magnitude of selection bias, and carry out MLE based on the full data to assess the extent of efficiency loss due to missing data. Results for estimation of the effect of exposure $\beta_1$ are summarized in table 1. The results for other coefficients $(\beta_0,\beta_2,\beta_3)$ follow a similar pattern as those included in table 1 and are therefore relegated to the supplementary materials. 

\begin{table}[!]
\begin{tabular}{c c | c c c c c| c c c c c}
\hline\noalign{\smallskip}
\multicolumn{2}{c|}{} & \multicolumn{5}{c|}{$n=1000$} & \multicolumn{5}{c}{$n=2000$} \\
\multicolumn{2}{c|}{Method}	& Bias & MCV & AV & ARE & \%Cover & Bias & MCV & AV & ARE & \%Cover \\
\hline
\multirow{2}{*}{UMLE} & IPW  &  0.01 & 0.11 & 0.10 & {} & 94.6 &  0.00 & 0.05 & 0.05 & {} & 96.1\\
& AIPW &  0.00 & 0.08 & 0.07 & 0.70 & 93.8  &  0.00 & 0.03 & 0.03 & 0.67 & 95.6\\
\hline
\multirow{2}{*}{CBE} & IPW & -0.03 & 0.11 & 0.10 & {} & 94.4 & -0.01 & 0.05 & 0.05 & {} & 94.8\\
& AIPW & -0.02 & 0.08 & 0.07 & 0.70  & 93.3 & 0.00 & 0.04 & 0.04 & 0.74 & 94.2 \\
\hline
\multicolumn{2}{c|}{CC}  & -0.51 & 0.13 & 0.12 & {} & 66.4 & -0.50 & 0.06 & 0.06 &{} & 46.6 \\
\multicolumn{2}{c|}{Full MLE} &  0.00 & 0.05 & 0.05 & {} & 94.4  &  0.00 & 0.02 & 0.03 &{}& 95.8 \\
\hline
\multicolumn{12}{l}{\footnotesize{Values of zero correspond to less than 0.005}}

\end{tabular}
\caption{Estimation for the effect of exposure $\beta_1=-0.4$ in the logistic regression model from 1000 simulation replicates. Results for UMLE are restricted to converged replicates (57.9\% and 64.0\% for $n=1000,2000$). MCV gives the Monte Carlo variance. AV refers to the estimate of asymptotic variance and ARE is the estimate of asymptotic relative efficiency comparing AIPW to IPW based on AV. Coverages are based on nominal 95\% Wald confidence intervals.}
\end{table}

Biases for IPW and AIPW estimators of $\beta_1$ using UMLE or CBE generally decrease with increasing sample size, and biases become negligible at moderate sample sizes. There is a slight downward bias of the asymptotic variance estimator compared to the Monte Carlo variance at $n=1000$, particularly for the AIPW estimator, leading to an empirical coverage slightly below the nominal 95\% level. However the performance of the asymptotic variance estimator and corresponding coverage improves at $n=2000$. The asymptotic relative efficiency (ARE) of AIPW compared to IPW is approximately $0.70$ based on estimated asymptotic variances, in agreement with theory. 

The CC estimator has substantial bias and poor coverage irrespective of sample size. Although the CC estimator may be unbiased when the missing data model only depends on regression covariates but not on the outcome, it is clearly biased under the current data generating mechanism which allows for arbitrary missing data patterns involving both the outcome and covariates. The CC estimator has larger standard errors compared to either the IPW or the AIPW estimator. 

The proportion of simulation replicates for which the UMLE converged increased slightly with a doubling of sample size (57.9\% and 64.0\% for $n=1000,2000$). Based on the derivative-free approximation to the UMLE solution, we calculated the smallest estimated complete case probability in each simulation replicate. These values for non-convergence cases hover around zero, suggesting that, as we have previously hypothesized, lack of convergence of the UMLE approach may be due mostly to empirical complete case probabilities that effectively violate the positivity assumption (\ref{eq:assume2}), which may occur by chance particularly in small samples, even when the assumption holds in the population. 

We note that the bias of the IPW or AIPW estimator using CBE for the missing data model is slightly larger than that using UMLE when the latter converges. Similar finite sample bias has been reported in previous implementations of the CBE in a log-linear model of risk \citep{chucole1}. However, even so, as noted above the coverage of $95\%$ confidence intervals does not appear to be affected and the finite sample bias appears to vanish relatively fast with increasing sample size. Furthermore, the BCE is guaranteed to produce an estimate for the complete case probabilities within the parameter space of the model which may be subsequently used for IPW or AIPW. 

\section{APPLICATION}
\label{sec:app}

The empirical application concerns a study of the association between maternal exposure to highly active antiretroviral therapy (HAART) during pregnancy and birth outcomes among HIV-infected women in Botswana. A detailed description of the study cohort has been presented elsewhere \citep{chen}. The entire study cohort consists of 33148 obstetrical records abstracted from 6 sites in Botswana for 24 months. Our current analysis focuses on the subset of women who were known to be HIV positive ($n=9711$). The birth outcome of interest is preterm delivery, defined as delivery $<37$ weeks gestation. 6.7\% of the outcomes are unobserved. The data also contains a number of predictors of interest with unobserved values (Table 2):  maternal hypertension in pregnancy (6.5\% missing), whether CD$4^+$ cell count is less than 200 $\mu$L (53.4\% missing) and whether a woman continued HAART from before pregnancy or not. Our goal is to correlate these factors with preterm delivery. We applied the proposed IPW and AIPW estimators in logistic regression as well as performed CC analysis. We also provide results for multivariate imputation by chained equations (MICE) \citep{buuren2, mice} as comparison (Table 3). 

\begin{table}[!]
\centering
\begin{tabular}{cccccr}
  \hline\noalign{\smallskip}
Pattern (R) & Preterm Delivery & Hypertension & Low CD$4^+$ & Cont. HAART & {\% of data} \\   
  \hline
1 &   1 &   1 &   1 &   1 & 43.7 \\ 
  2 & 0 &   1 &   1 &   1 & 2.0 \\ 
  3 &   1 & 0 &   1 &   1 & 0.7 \\ 
   4 &   0 & 0 &   1 &   1 & 0.2 \\ 
  5 &   1 &   1 & 0 &   1 & 44.9 \\ 
  6 & 0 &   1 & 0 &   1 & 2.9 \\ 
  7 &   1 & 0 & 0 &   1 & 4.0 \\ 
  8 & 0 & 0 & 0 &   1 & 1.6 \\ 
   \hline
\end{tabular}
\caption{Tabulation of non-monotone missing data patterns as a percentage of total data ($n=9711$). Missing variables are indicated by 0. Complete-cases are given in the first pattern.}
\end{table}

\begin{table}[!]
\centering
\begin{tabular}{c c c c}
\hline\noalign{\smallskip}

{Method}                     & {Hypertension} &{Low CD$4^+$} & {Cont. HAART} \\
\hline
{CC}                            &  1.29 (1.06, 1.57) & 1.12 (0.89, 1.40) & 1.31 (1.04, 1.65) \\
{IPW}                          &  1.55 (1.20, 2.01) &  1.12 (0.84, 1.50)&  1.53 (1.18, 1.99) \\
{AIPW}                        &  1.41 (1.23,  1.62) & 1.08 (0.88, 1.34) & 1.47 (1.29, 1.66)\\
{MICE}                         & 1.34 (1.17, 1.54)  & 1.03 (0.77, 1.39) & 1.22 (1.09, 1.36)\\
\hline
\multicolumn{4}{c}{Analysis combining missing data patterns $R=3,4$}\\
\hline
{IPW}                          &  1.55 (1.20, 2.01) &  1.13 (0.85, 1.52)&  1.52 (1.18, 1.97)\\
{AIPW}                        &  1.40 (1.22,  1.61) & 1.11 (0.90, 1.37) & 1.46 (1.28, 1.66)\\
\hline
\end{tabular}
\caption{Analysis for outcome preterm delivery with estimated odds ratios from logistic regression. Wald 95\% confidence intervals for IPW /AIPW estimators are based on estimated asymptotic variances. The standard error for MICE is estimated by Rubin's formula \citep{rubin3} with $M=50$ imputed samples.}
\end{table}

The IPW estimator of the logic model for preterm delivery uses to estimate the weight a missing data model of the form given by (\ref{eq:logit}), which includes only the main effects of observed variables $L_{(m)}$ for each missing data pattern $m=2,...,8$.  The UMLE converged in this dataset. Given the fairly large sample size $(n=9711)$, the results for IPW are similar using UMLE and CBE to estimate the missing data process, consistent with findings from both the simulation study and asymptotic theory. Hence, only results for CBE are presented for the IPW estimator in Table 3. AIPW estimator is implemented as outlined in the simulation study. MICE specifies a univariate imputation model for each of the incomplete variables preterm delivery, maternal hypertension and low CD$4^+$ (the variable continued HAART treatment is fully observed in the sample and not imputed). The binary variables preterm delivery, hypertension and low CD$4^+$ are imputed using logistic regressions, to provide a total of $M=50$ imputed data sets for linear regression before pooling the results in the final analysis. In a separate analysis, the two sparsely observed missing data patterns $R=3,4$ with 75 and 15 samples respectively are combined into one pattern. The probability of observing this combined pattern depends on the set of covariates $L_{(3)} \cap L_{(4)}$, i.e.  low CD$4^+$ and continued HAART treatment.

The IPW and AIPW estimated odds ratio for preterm delivery associated with maternal hypertension and continued HAART treatment increased by approximately $15\%$ respectively compared to CC estimates. The point estimates of the effect for low CD$4^+$ are similar between CC and IPW. The observed ARE of AIPW compared to IPW differs across different coefficients: $0.28$ for maternal hypertension, $0.53$ for low CD$4^+$ and $0.24$ for continued HAART treatment. The observed ARE of AIPW compared to MICE are $1.00$ for maternal hypertension, $0.51$ for low CD$4^+$ and $1.25$ for continued HAART treatment. Point estimates from MICE show that the odds ratio for preterm delivery associated with maternal hypertension increased marginally by about $4\%$, but the odds ratios associated with low CD$4^+$ and continued HAART treatment decreased by $8\%$ and $7\%$ respectively. The analysis which combines missing data patterns $R=3,4$ for IPW/AIPW gives similar results to the original analysis.

Differences between MICE and IPW / AIPW estimates may reflect differences of modeling assumptions since the former relies on model assumptions about full data univariate conditional laws while the latter relies on a model for the missing data mechanism. In the current application, neither the conditional distribution of covariates in the full data nor the missing data model is of primary scientific interest. Although model compatibility of the conditional laws specified in MICE may be an issue \citep{white, buuren}, simulation studies suggest that this may not be a serious problem in practice \citep{buuren3}. In general, more efficient estimators can be obtained by specifying a full data model. However, in this illustration, the proposed AIPW estimator yields efficiency gains relative to IPW comparable to MICE, while at the same time entirely avoiding the need to model the full data law.

\section{DISCUSSION}
\label{sec:discuss}

We have proposed a simple yet general class of missing data models for nonmonotone MAR mechanisms which makes no assumption about the full data distribution. Our models are explicit in their dependence on only the observed variables, and the proposed IPW estimator can easily be implemented using existing software. The paper makes two important contributions, first we describe a simple UMLE approach to estimate the missing data mechanism that is straightforward to implement although that may suffer from convergence issues in small samples. Our second contribution offers a remedy to failure of UMLE by introducing a constrained Bayesian estimator which circumvents any potential convergence difficulty encountered with UMLE. Another contribution shows that AIPW can achieve substantial gains in efficiency over simple IPW estimators by recovering information from incomplete cases, while avoiding having to model the full data distribution.

Assuming no model misspecification, the proposed IPW / AIPW estimators corrects the bias of CC analysis and may be used whenever one has available a full data estimating equation and the nonmonotone MAR missing data mechanism potentially depends also on the outcome. While CBE is guaranteed to produce valid probability weights for subsequent estimation of a full data regression or other functionals of interest, we found that there may be some finite sample bias in small to moderate samples. However, this bias appears to vanish with increasing sample size. The bias may be due to the fact that constraints (\ref{eq:cons2}) are imposed on complete-cases only, and thus the constraints may not be satisfied for the incomplete cases. The CBE approach could be adapted to impose the constraints over a finite range of possible values for the full data $L$, if bounds for the sample space were known.

Lastly, Robins and Gill have argued that the class of RMM models represents the most general plausible physical mechanism for generating non-monotone missing data \citep{jamie1}. Therefore, they have effectively argued that any model within our class that is not RMM may be difficult to motivate scientifically. We emphasize that the perspective we have presented is completely agnostic as to whether a particular submodel of MAR may be more scientifically meaningful than another; in fact, RMM, like any other submodel of MAR, can be accommodated by the proposed approach, but would require placing additional constraints while sampling from the posterior, to ensure that one remains within the submodel. This will necessarily result in a more complicated fitting procedure, with little apparent benefit for bias reduction or efficiency gain. This is because, as well established in the missing data literature, it is generally advisable for efficiency considerations in IPW estimation under MAR, that one estimates the probability of a complete-case using as richly parameterized a regression as empirically feasible \citep{jamie2}. This implies that even if RMM is correctly specified, one would generally benefit from including correlates of the full data estimating equation into a model for the missing data mechanism, even if such variables do not necessarily correlate with the missing data process. We believe such efficiency considerations trump any concern for scientific interpretation of the model for the missing data process, particularly since after all, the missing data process is technically a nuisance parameter not of primary scientific interest. 

\bigskip
\begin{center}
{\large\bf APPENDIX: PROOFS}
\end{center}

\subsection*{A.1 Restrictions imposed by polytomous logistic regression model}
Suppose there are $M$ missingness patterns, each with observed variables $L_{(m)}$, $m=1,...,M$. Choosing pattern $j$ as the baseline category, we model the other missingness pattern probabilities as 
\begin{align*}
\Pr \{R=m|L\}= \frac{\exp({\gamma_m}^{\prime}L_{(m)} )}{1+\sum_{k\in \{1,...,M\} \setminus \{j\}}  \exp({\gamma_k}^{\prime}L_{(k)} ) } \phantom{-} \text{for} \phantom{-} m \in \{1,...,M\} \setminus \{j\}.
\end{align*}
Let $L_I=\bigcap_{m \in \{1,...,M\} \setminus \{j\}} L_{(m)}$. Then by the MAR assumption, each of the above probabilities $\Pr \{R=m|L\}$ depends on $L_{(m)}$ respectively. But they can only depend on $L_I$.  If not, then the probability for one of the missing data patterns $h$ will depend on variables $L_{(h)} \setminus L_I$ that another pattern does not have. This is not possible due to the linked nature of the terms in the denominator of the probability expression.
\subsection*{A.2 Asymptotic results for IPW estimator}

The consistency of $\hat{\beta}$ can be established under general conditions for 2-step estimators \citep{newey} to show uniform convergence of estimating equation  (\ref{eq:ipweqn}) in $\beta$, where we make use of the fact that $\hat{\bg} \overset{p}{\to} \bg$. Typically one would need to impose moment assumptions on $\pi_1 (L;\gamma)$ and $M(L;\beta)$ \citep{wooldridge}.

To investigate the asymptotic distribution of $\hat{\beta}$, under suitable regularity conditions expand  (\ref{eq:ipweqn}) around the true values $\beta_0$ and subsequently $\gamma_0$, 
\begin{align*}
&\sqrt{n}(\hat{\beta}-\beta_0) =- \left[\frac{1}{n}\sum_{i=1}^{n} \nabla_\beta \Gamma_i(\beta^*,\hat{\gamma}) \right]^{-1}\frac{1}{\sqrt{n}}\sum_{i=1}^{n}  \Gamma_i(\beta_0,\hat{\gamma}) \\
&=- \left[\frac{1}{n}\sum_{i=1}^{n} \nabla_\beta \Gamma_i(\beta^*,\hat{\gamma})  \right]^{-1} \times \left[ \frac{1}{\sqrt{n}}\sum_{i=1}^{n}  \Gamma_i(\beta_0,\gamma_0) +\left( \frac{1}{n}\sum_{i=1}^{n} \nabla_{\gamma}\Gamma_i(\beta_0,\gamma^*)  \right)\sqrt{n}(\hat{\gamma}-\gamma_0) \right]
\end{align*}
where $\beta^*$ and $\gamma^*$ are the mean values and $ \Gamma(\beta,\gamma) = \{ \mathds{1}(R_{1}=1)/\pi_1 (L;\gamma)\}M(L;\beta)$. When $\hat{\gamma}$ is the maximum likelihood estimator or a Bayes point estimator satisfying conditions in the Bernstein-von Mises Theorem, it is an asymptotically linear estimator with the influence function
\begin{align}
\sqrt{n}(\hat{\gamma}-\gamma_0)=\frac{1}{\sqrt{n}} \sum_{i=1}^{n} \mathrm{E} \left[S_{\gamma_0} S_{\gamma_0}^{T}\right]^{-1} S_i{\gamma_0}+o_p(1)
\end{align}
where $S_{\gamma}$ is the score function with respect to the missing data model parameters $\gamma$. Substituting the influence function representation into previous expansion gives
\begin{align}
&\sqrt{n}(\hat{\beta}-\beta_0) \notag \\
&=-\mathrm{E}\{\nabla_\beta \Gamma(\beta_0,\gamma_0) \}^{-1}\frac{1}{\sqrt{n}} \sum_{i=1}^{n} \left\{ \Gamma_i(\beta_0,\gamma_0) +\mathrm{E}\{\nabla_\gamma \Gamma(\beta_0,\gamma_0) \}\mathrm{E} \left[S_{\gamma_0} S_{\gamma_0}^{T}\right]^{-1} S_i{\gamma_0} \right\} +o_p(1).
\end{align}
In addition, from the assumption that the parameters governing full data and the missing data process are separable, under standard regularity conditions we have for observed data $\bo$
\begin{align*}
&\mathrm{E} [\Gamma(\beta,\gamma)] = \int \Gamma(\beta,\gamma) f(\bo;\beta,\gamma) \,d\bo =0 \\
&\frac{\partial}{\partial \gamma}\mathrm{E} [\Gamma(\beta,\gamma)] = 
 \int \frac{\partial}{\partial \gamma}\Gamma(\beta,\gamma) f(\bo;\beta,\gamma) \,d\bo +  \int \Gamma(\beta,\gamma) \frac{\partial}{\partial \gamma} f(\bo;\beta,\gamma) \,d\bo =0 \\
&\implies \mathrm{E}\{\nabla_\gamma \Gamma(\gamma,\beta)\}=-\int \Gamma(\beta,\gamma) \frac{\frac{\partial}{\partial \gamma} f(\bo;\beta,\gamma)}{ f(\bo;\beta,\gamma)} f(\bo;\beta,\gamma) \,d\bo  = -\mathrm{E} [\Gamma (\beta,\gamma)S_{\gamma}].
\end{align*}
Substituting the above equality to (27)
\begin{align*}
&\sqrt{n}(\hat{\beta}-\beta_0)= \notag \\
& -\mathrm{E}\{\nabla_\beta \Gamma(\beta_0,\gamma_0) \}^{-1}\frac{1}{\sqrt{n}} \sum_{i=1}^{n} \left\{ \Gamma_i(\beta_0,\gamma_0)-\mathrm{E} [\Gamma (\beta_0,\gamma_0)S_{\gamma_0}^T]\mathrm{E} \left[S_{\gamma_0} S_{\gamma_0}^{T}\right]^{-1} S_i{\gamma_0} \right\} +o_p(1) \notag.
\end{align*}
An application of Slutsky's theorem shows that
\begin{align}
\sqrt{n}(\hat{\beta}-\beta_0) \xrightarrow[]{d} N\left(0, \mathrm{E}\{\nabla_\beta \Gamma(\beta_0,\gamma_0) \}^{-1}\mathrm{Var}\left[  \Gamma(\beta_0,\gamma_0)-W(\beta_0,\gamma_0)\right]\mathrm{E}\{\nabla_\beta \Gamma(\beta_0,\gamma_0) \}^{-1^T}\right)
\end{align}
 where
$$
W(\beta_0,\gamma_0)=\mathrm{E} [\Gamma (\beta_0,\gamma_0)S_{\gamma_0}^T]\mathrm{E} \left[S_{\gamma_0} S_{\gamma_0}^{T}\right]^{-1} S{\gamma_0}.
$$
The sandwich estimator is consistent for $\mathrm{E}\{\nabla_\beta \Gamma(\beta_0,\gamma_0) \}^{-1}\mathrm{E}\left[ \Gamma(\beta_0,\gamma_0)^{\bigotimes2}\right] \mathrm{E}\{\nabla_\beta \Gamma(\beta_0,\gamma_0) \}^{-1^T}$. In the Hilbert space of mean-zero random functions, $\mathrm{E} \left[\Gamma (\beta_0,\gamma_0)S_{\gamma_0}^T\right]\mathrm{E} \left[S_{\gamma_0} S_{\gamma_0}^{T}\right]^{-1} S{\gamma_0}$ is the projection of $\Gamma(\beta_0,\gamma_0)$ onto the linear subspace spanned by elements of $S_{\gamma_0}$. Therefore by Pythagorean Theorem 
$$
\mathrm{E}\left[ \Gamma(\beta_0,\gamma_0)^{\bigotimes2}\right]-\mathrm{E}\left[ \left\{ \Gamma(\beta_0,\gamma_0) -\mathrm{E} \left[\Gamma (\beta_0,\gamma_0)S_{\gamma_0}^T\right]\mathrm{E} \left[S_{\gamma_0} S_{\gamma_0}^{T}\right]^{-1} S{\gamma_0}\right\}^{\bigotimes2}\right]\
$$ 
is positive semi-definite and the sandwich estimator provides conservative estimate for the true asymptotic variance.

\bibliographystyle{agsm}

\bibliography{nmmbib}

\bigskip
\begin{center}
{\large\bf SUPPLEMENTARY MATERIALS}
\end{center}

\begin{table}[!htbp]
\begin{tabular}{c c | c c c c c| c c c c c}
\hline\noalign{\smallskip}
\multicolumn{2}{c}{} & \multicolumn{5}{c}{$n=1000$} & \multicolumn{5}{c}{$n=2000$} \\
\multicolumn{2}{c|}{Method} & Bias & MCV & AV &ARE & \%Cover & Bias & MCV & AV & ARE & \%Cover \\
\hline
\multicolumn{10}{c}{$\beta_0$}\\
\hline
\multirow{2}{*}{UMLE} & IPW  &  0.01 & 0.08 & 0.08 & {} & 95.0 &  0.00 & 0.04 & 0.04 & {} & 94.0\\
& AIPW  &  0.00 & 0.07 & 0.06 &0.75 & 94.6 &  -0.01 & 0.03 & 0.03 & 0.75 & 93.3\\
\hline
\multirow{2}{*}{BCE} & IPW & 0.08 & 0.08 & 0.08 & {} & 95.4 & 0.04 & 0.04 & 0.04 & {} & 94.8\\
& AIPW& 0.03 & 0.07 & 0.07  & 0.80 & 95.1  & 0.01 & 0.03 & 0.03  & {} & 94.1 \\
\hline
\multicolumn{2}{c|}{CC} & 1.35 & 0.11 & 0.11 & {} & 1.8  & 1.33 & 0.05 & 0.06 & 0.75 & 0.0 \\
\multicolumn{2}{c|}{MLE} &  0.01 & 0.04 & 0.04 & {} & 95.0  &  0.00 & 0.02 & 0.02 & {} & 95.6 \\
\hline
\multicolumn{10}{c}{$\beta_2$}\\
\hline
\multirow{2}{*}{UMLE} & IPW  &  0.00 & 0.10 & 0.10 & {} & 94.5 &  0.00 & 0.05 & 0.05 &{} & 96.7\\
& AIPW &  0.00 & 0.09 & 0.08 & 0.93 & 94.0  &  0.00 & 0.04 & 0.04 & 0.82 & 95.8\\
\hline
\multirow{2}{*}{BCE} & IPW& -0.03 & 0.10 & 0.10 & {} &  95.3  & -0.02 & 0.05 & 0.05 & {} & 96.1\\
& AIPW & -0.02 & 0.09 & 0.08  & 0.93 & 94.3 & -0.01 & 0.04 & 0.04  & 0.82 & 94.9 \\
\hline
\multicolumn{2}{c|}{CC}  & -0.25 & 0.12 & 0.12 &  {} & 88.5 & -0.25 & 0.06 & 0.06 & {} & 82.5 \\
\multicolumn{2}{c|}{MLE}  &  -0.01 & 0.05 & 0.05 & {} & 94.5 &  0.00 & 0.02 & 0.02 & {} & 94.5 \\
\hline
\multicolumn{10}{c}{$\beta_3$}\\
\hline
\multirow{2}{*}{UMLE} & IPW  &  -0.02 & 0.11 & 0.10 & {} & 94.3 &  0.00 & 0.05 & 0.05 &{} & 94.5\\
& AIPW&  -0.01 & 0.10 & 0.08 & 0.82 & 94.5  &  0.01 & 0.04 & 0.04 & 0.79 & 93.3 \\
\hline
\multirow{2}{*}{BCE} & IPW  & -0.05 & 0.11 & 0.11 &{} & 94.4 & -0.02 & 0.05 & 0.05 & {} & 95.0\\
& AIPW & -0.01 & 0.10 & 0.08  & 0.77 & 93.8 & 0.00 & 0.04 & 0.04  & 0.79 & 93.4 \\
\hline
\multicolumn{2}{c|}{CC} & -0.47 & 0.12 & 0.12 & {} & 74.2  & -0.44 & 0.06 & 0.06 & {} & 53.4 \\
\multicolumn{2}{c|}{MLE}  &  -0.01 & 0.05 & 0.05 & {} & 94.6  &  0.00 & 0.02 & 0.02 & {} & 95.7\\
\hline
\multicolumn{12}{l}{\footnotesize{Values of zero correspond to less than 0.005}}
\end{tabular}
\caption{Estimation for the effect of exposure $\beta_1=-0.4$ in the logistic regression model from 1000 simulation replicates. Results for UMLE are restricted to converged replicates (57.9\% and 64.0\% for $n=1000,2000$). MCV gives the Monte Carlo variance. AV refers to the estimate of asymptotic variance and ARE is the estimate of asymptotic relative efficiency comparing AIPW to IPW based on AV. Coverages are based on nominal 95\% Wald confidence intervals.}
\end{table}

\end{document}